%% file: main.tex
\documentclass{INTERSPEECH2023}
\usepackage{graphicx}
\usepackage{caption}
\usepackage{subcaption}
\usepackage{pgfplots, mathtools}
\usepackage{xcolor}
\usepackage{tikz}
\usepackage{pdflscape}
\usetikzlibrary{matrix}
\usepackage{cite}
\input{preamble}
\interspeechcameraready
\UseRawInputEncoding

\title{HD-DEMUCS: General Speech Restoration with \\ Heterogeneous Decoders}

\name{Doyeon Kim$^1$, Soo-Whan Chung$^2$, Hyewon Han$^1$, Youna Ji$^2$, Hong-Goo Kang$^1$}
\address{
  $^1$Department of Electrical and Electronic Engineering, Yonsei University, South Korea\\
  $^2$NAVER Cloud, South Korea}
\email{ehyeon24@dsp.yonsei.ac.kr, soowhan.chung@navercorp.com, hwhan@dsp.yonsei.ac.kr, youna.ji@navercorp.com, hgkang@yonsei.ac.kr}

\begin{document}

\maketitle



\input{1_abstraction}
\input{2_introduction}
\input{3_related}
\input{4_proposed}
\input{5_experiments}
\input{6_conclusion}

\clearpage
\label{section:references}

\bibliographystyle{IEEEtran}
\bibliography{shortstrings,mybib}

\clearpage

\end{document}

%% file: preamble.tex
\newcommand{\shortname}{HD-DEMUCS}
\newcommand{\fullname}{Heterogeneous Decoders-DEMUCS}
\newcommand{\encoder}{$\mathbf{E}$}
\newcommand{\suppression}{$\mathbf{D_s}$}
\newcommand{\refinement}{$\mathbf{D_r}$}
\newcommand{\fusion}{$\mathbf{F}$}
\newcommand{\skipconnect}{$\mathbf{S_{\text{\suppression}\rightarrow \text{\refinement}}}$}

\definecolor{turquoise}{cmyk}{0.65,0,0.1,0.3}
\definecolor{purple}{rgb}{0.65,0,0.65}
\definecolor{dark_green}{rgb}{0, 0.5, 0}
\definecolor{orange}{rgb}{0.8, 0.6, 0.2}
\definecolor{red}{rgb}{0.8, 0.2, 0.2}
\definecolor{darkred}{rgb}{0.6, 0.1, 0.05}
\definecolor{blueish}{rgb}{0.0, 0.3, .6}
\definecolor{light_gray}{rgb}{0.7, 0.7, .7}
\definecolor{pink}{rgb}{1, 0, 1}
\definecolor{greyblue}{rgb}{0.25, 0.25, 1}
\definecolor{dark_gray}{rgb}{0.3, 0.3, 0.3}

\newcommand{\tblue}[1]{{\color{blueish}#1}}
\newcommand{\tred}[1]{{\color{darkred}#1}}

\newcommand{\tgray}[1]{{\color{dark_gray}#1}}
\newcommand{\tgreen}[1]{{\color{dark_green}#1}}

\newcommand{\Figure}[1]{Figure~\ref{fig:#1}}

\newcommand{\Table}[1]{Table~\ref{tab:#1}}

\renewcommand{\paragraph}[1]{\noindent\textbf{#1}.}

\newcommand{\eg}{\textit{e.g.,~}}
\newcommand{\ie}{\textit{i.e.,~}}

%% file: 1_abstraction.tex
\begin{abstract}
This paper introduces an end-to-end neural speech restoration model, \shortname, demonstrating efficacy across multiple distortion environments.
Unlike conventional approaches that employ cascading frameworks to remove undesirable noise first and then restore missing signal components, our model performs these tasks in parallel using two heterogeneous decoder networks.
Based on the U-Net style encoder-decoder framework, we attach an additional decoder so that each decoder network performs noise suppression or restoration separately.
We carefully design each decoder architecture to operate appropriately depending on its objectives.
Additionally, we improve performance by leveraging a learnable weighting factor, aggregating the two decoder output waveforms. 
Experimental results with objective metrics across various environments clearly demonstrate the effectiveness of our approach over a single decoder or multi-stage systems for general speech restoration task.

\end{abstract}

%% file: 2_introduction.tex
\section{Introduction}

Speech signal is a fundamental and intuitive medium for human interaction. 
However, various distortions are often present in observed speech, including background noise, reverberation, and cross-talk from other speakers. 
These distortions can severely degrade the perceptual quality of input signals, posing challenges for understanding the target speech.
Additionally, acoustic responses such as room impulse response and transmission channel distortions can alter the spectral composition of speech signal, resulting in poor clarity and intelligibility. 
To address these issues, speech enhancement has become a crucial pre-processing step that aims to improve the perceptual quality and intelligibility of input speech by mitigating undesirable distortion effects. 
By enhancing speech signals, speech-based applications, such as automatic speech recognition~\cite{schneider2019wav2vec, baevski2020wav2vec, suyoun2017ctc} and speaker verification~\cite{Desplanques2020ecapa, Rui2022efficienttdnn, mirco2018sincnet}, can provide more accurate and reliable results, leading to improved user experiences.

Recent deep learning-based methods have shown remarkable performance in speech enhancement, primarily by reducing noise and reverberation~\cite{tagliasacchi2020seanet, kim2020tgsa, fu2022metricganu}. 
In~\cite{fu2019metricgan, fu2021metricgan+, lee2019tfmask}, the authors have predicted a spectrogram or spectral mask to suppress distortions. 
In~\cite{Kegler2020deep,borsos2022speechpainter, moliner2022behm}, they have attempted to generate missing components, including spectral bands or temporal occlusions, leveraging the impressive predictive capability of neural networks.
Furthermore, some works have produced more realistic speech from distorted inputs by introducing generative models such as generative adversarial network~\cite{pascual2017segan,su2020hifi} and diffusion-based score-matching method~\cite{lu2022conditional,welker2022speech}.

In real-world scenarios, speech degradation factors do not occur in isolation but rather in correlation with each other, incurring challenges in speech enhancement tasks.
However, most speech enhancement methods have traditionally focused on processing a single distortion and have dealt with multiple distortions by cascading several task-oriented models~\cite{liu2022voicefixer, strake2019separated},  neglecting correlations between various distortions.
This fact raises concerns that artifacts (\eg musical noise, remaining distortions) caused by the front-end speech enhancement method are propagated downstream, resulting in severe degradation of post-enhancement modules.
In~\cite{liu2022voicefixer}, the authors have defined the task of handling multiple distortions as \textit{general speech restoration}, which refers to speech restoration task in this paper, solving the problem by training neural enhancement models with adversarial training.
They have designed their methods based on the analysis-and-synthesis point of view, \ie restoring mel-spectrograms by a residual U-Net structure~\cite{Kong2021DecouplingMA} and generating waveforms from mel-spectrograms using an extra vocoder.
In~\cite{lemercier2022analysing}, the authors have introduced a generative diffusion-based method that produces high-quality speech waveforms from distorted inputs, beyond eliminating complex distortions.
Although various authors have exhibited impressive restoration performance, further improvements are possible by designing a neural network that considers the characteristics of various distortion types present in the input signals.

In this paper, we propose a novel end-to-end speech restoration network, \fullname~(\shortname).
Unlike traditional methods that combine separate models to address different restoration tasks, our model achieves improved efficiency with two parallel decoder networks.
Our approach leverages the well-known encoder-decoder framework, DEMUCS~\cite{defossez2020real}, which has demonstrated its effectiveness in suppressing noise and reverberation.
The novelty of \shortname~mainly comes from the modification of the decoder network, it includes two heterogeneous decoders that are designed to perform different restoration tasks efficiently.
Specifically, one decoder, a \textit{suppression} decoder, focuses on distortion removal by suppressing additive and convolution distortions, rather than generating clean speech.
In contrast, another decoder, a \textit{refinement} decoder, is responsible for generating clean speech with fine perceptual quality, by restoring missing components on input speech.
They collaborate by providing latent features from the suppression to the refinement decoder, as the refinement process can be addressed more efficiently with enhanced features rather than solely encoded features.
Additionally, we customize the configuration of the decoders based on their respective restoration tasks.
The final restored speech waveforms are obtained by summing the outputs of the two decoders through a fusion module. 
 Our experiments and ablation studies demonstrate the effectiveness of our proposed method and highlight the importance of the submodules in processing multiple distortions simultaneously.

%% file: 3_related.tex
\input{FigTex/overall}

\vspace{-5pt}
\section{DEMUCS}
\label{section:related}

The most relevant work to ours is DEMUCS, which was proposed for 
speech enhancement task using a U-Net-based encoder-decoder architecture.
The encoder receives upsampled time-domain distorted input speech and analyzes it through the stack of convolutional blocks, producing a latent embedding.
The encoder and the decoder each have five convolution and de-convolution blocks.
They benefit from the large receptive fields of strided convolution layers, resulting in improved contextual analysis and representation capability, followed by Gated Linear Units (GLUs) \cite{Dauphin2017glu}.
Additionally, a Long Short-Term Memory (LSTM) layer between the encoder and decoder strengthens the sequential modeling that cannot be achieved in the encoder convolution layers. 
The encoder and decoder blocks are connected using U-Net skip connections~\cite{olaf2015unet} to preserve information during network propagation.
The final enhanced speech is obtained by downsampling the decoder output and multiplying it by the standard deviation of the input speech.

%% file: FigTex/overall.tex
\begin{figure*}[t]
  \centering
  \includegraphics[width=0.8\linewidth]{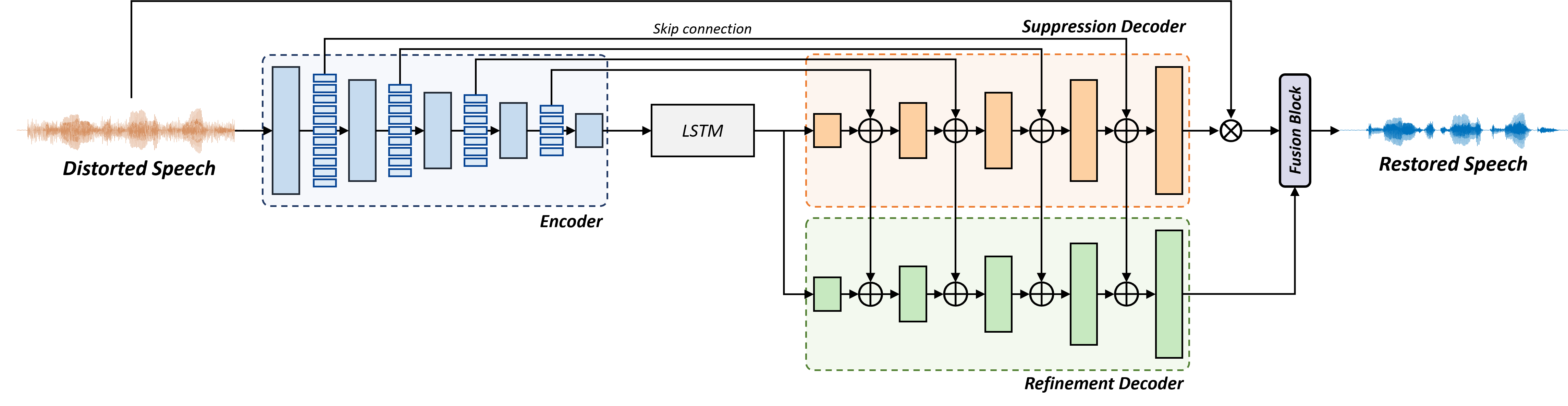}
    \vspace{-5pt}
  \footnotesize{
  \caption{Illustration of the proposed speech restoration network, \shortname.}}
  \label{fig:overall}
  \vspace{-10 pt}
\end{figure*}

%% file: 4_proposed.tex
\input{FigTex/submodules}

\section{Proposed model} 
\label{section:proposed}
\subsection{Problem formulation}
In this paper, we aim to restore speech in cases where input speech includes background noise, reverberation, and frequency band distortion.
Here, we can formulate the input speech $y$ as:
\vspace{-3pt}
\begin{equation}
    y = h(x*r) + n,
    \vspace{-3pt}
\end{equation}
where $x$ and $n$ are a clean speech and background noise respectively, $*$ is the convolution symbol, and $r$ reflects the room impulse response.
Specifically, $h$ creates spectral distortions, modeled by a high-pass, low-pass, and band-pass filter.

\input{tables/baselines}

\subsection{Overall architecture}
We propose a novel end-to-end speech enhancement network, \shortname, which simultaneously processes multiple distortions of diverse characteristics in input speech.
We design the overall architecture with an analysis and synthesis approach; thus, we focus more on the functionality of decoders rather than that of the encoder which takes the analysis stage.
Therefore, we have integrated a new decoder framework to handle the encoded representations.
We categorize the speech restoration task into two perspectives: \textit{suppression} and \textit{refinement}.
Following these perspectives, we allocate two heterogeneous decoders to address distinct distortion types.
Figure 1 illustrates the overall architecture of \shortname, comprising 4 submodules: an encoder \encoder, a suppression decoder \suppression, a refinement decoder \refinement, and a fusion block \fusion.

\paragraph{Encoder}
The encoder follows the well-designed architecture of the causal DEMUCS with convolution layers of 48 hidden channels and an LSTM layer.
The encoder block consists of five blocks, and each block consists of the convolutional layer with a kernel size of 8 and stride of 4 followed by the GLU activation function.
Before the encoder block, an upsampling layer increases the sampling rate by a factor of 4.
The encoder provides its intermediate embeddings to the suppression decoder with skip connections ($\mathbf{S_{\text{\encoder} \rightarrow \text{\suppression}}}$).
Furthermore, the suppression decoder provides its latent embeddings, summed up with the skip connection, to the refinement decoder (\skipconnect).

\paragraph{Suppression decoder}
The objective of the suppression decoder is to eliminate unwanted additive or convolutive distortions, such as noise and reverberation tails, from the speech signal.
To accomplish this, the decoder receives the latent embedding of the encoder and computes a time-domain mask to suppress the input signal distortions rather than estimating the enhanced speech directly.
There exist five suppression blocks with skip connections from the encoder to prevent information leakage.
Each block of the suppression decoder, as illustrated in  \Figure{blocks}(\subref{fig:suppression_block}), consists of a series of 1-dimensional strided convolutional layers followed by the GLU function and transposed convolutional layers followed by the sigmoid activation function.
The kernel size and stride of the suppression block are identical to those of the encoder.
The output of the suppression decoder, where the range is limited between 0 and 1, is multiplied with the input signal, suppressing unwanted components.

\paragraph{Refinement decoder}
The purpose of the refinement decoder is to improve the perceptual quality and intelligibility of speech signals by refining or generating missing components.
Therefore, the output of the refinement decoder is a time-domain speech signal, in contrast to the suppression decoder which estimates the mask.
The refinement decoder in \Figure{blocks}(\subref{fig:generation_block}) utilizes two representations, one from the encoder output and the other from the intermediate representations of the suppression decoder.
Compared to the encoder output, the intermediate representation is expected to contain more refined information on additive distortions, which improves the refinement task efficiency.
Although the architectural composition of the refinement block is similar to that of the suppression block, there poses a critical difference in the transposed convolution layer.
The dilation factor in the transposed convolution layer is set to a value greater than one to increase the receptive field, as motivated by~\cite{vandenoord2016wavenet} for bandwidth extension.
We used a dilation factor of (1, 3, 5, 7, 9) on the layers of each block, with the kernel size and stride set to match those of the encoder.
It allows the refinement decoder to effectively enlarge the contextual information and estimate missing components caused by the distortions.

\paragraph{Fusion block}
To integrate the outputs of each decoder, we utilize a fusion block instead of simply adding the two decoder outputs.
The fusion block, inspired by~\cite{lai2023mixed}, employs a learnable weight to scale the outputs of the decoders for an effective combination.
As depicted in \Figure{blocks}(\subref{fig:fusion_block}), the two decoder outputs are stacked in the channel axis and passed through three convolutional layers.
Each convolution layer has a kernel size of 3 with stride of 1.
The fusion block employs a LeakyReLU activation function and a Sigmoid output function, to constrain the weight $w$ to a value lower than 1.
The suppression and refinement decoder outputs are scaled by $w$ and $(1-w)$ respectively, and then combined to produce the fusion block output as follows:
    \vspace{-3pt}
\begin{equation}
    \hat{x}_{up} = w\text{\refinement}(y_{up}) + (1-w)y_{up}\text{\suppression}(y_{up}),
    \vspace{-3pt}
\end{equation}
where $\hat{x}_{up}$ and $y_{up}$ indicate the (upsampled) restored and distorted speech signal, respectively.
Subsequently, the output is downsampled by an equivalent amount as the upsampling performed before the encoder, as in DEMUCS.

\subsection{Training criterion}
For a fair comparison with baseline methods, we adopted the same training criteria as in DEMUCS~\cite{defossez2020real}.
Both our proposed model and DEMUCS are trained by minimizing the distance between the estimated speech $\hat{x}$ and the reference speech $x$ in both the time and frequency domains.
In the time domain, we minimize the Euclidean distance between waveforms as follows:
    \vspace{-3pt}
\begin{equation}
    \mathcal{L}_T = \lVert x - \Hat{x}  \rVert_1,
    \vspace{-3pt}
\end{equation}
For the frequency domain, we employ a multi-resolution short-time Fourier Transform (MR-STFT) loss~\cite{Sercan2019fast, ryuichi2020parallel}.
First, the estimated and reference waveforms are transformed into magnitude spectra using various STFT configurations.
Then, we minimize the distance between spectra by considering both spectral convergence loss ($\mathcal{L}_{sc}$) and log-magnitude loss ($\mathcal{L}_{mag}$) for each STFT resolution.
The training loss on the frequency domain can be formulated
as below: 
    \vspace{-3pt}
\begin{equation}
    \mathcal{L}_F = \sum_{i=1}^M \left(\mathcal{L}^i_{sc} + \frac{1}{T}\mathcal{L}^i_{mag}\right),
    \vspace{-3pt}
\end{equation}
where $M$ is the number of STFT configurations, and $T$ defines the length of the speech.
For each resolution, $\mathcal{L}_{sc}$ and $\mathcal{L}_{mag}$ are defined as:
    \vspace{-3pt}
\begin{equation}
    \mathcal{L}_{sc} = {\lVert \mathbf{X} - \mathbf{\Hat{X}} \rVert}_F/{\lVert \mathbf{X} \rVert}_F,
    \vspace{-3pt}
\end{equation}
    \vspace{-3pt}
\begin{equation}
    \mathcal{L}_{mag} = \lVert \log\mathbf{X} - \log\mathbf{\Hat{X}} \rVert_1,    
    \vspace{-3pt}
\end{equation}
where $\mathbf{X}$ and $\mathbf{\Hat{X}}$ are the magnitude spectra of $x$ and $\Hat{x}$, $\lVert\cdot\rVert_F$ is Frobenius norm. 
We utilize three different configurations for the STFT, with the following parameters: number of FFT bins of (512, 1024, 2048), hop size of (50, 120, 240), and window length of (240, 600, 1200).

%% file: FigTex/submodules.tex
\begin{figure}[t]
  \centering
  \begin{subfigure}[b]{0.3\linewidth}
    \centering
    \includegraphics[width=\textwidth]{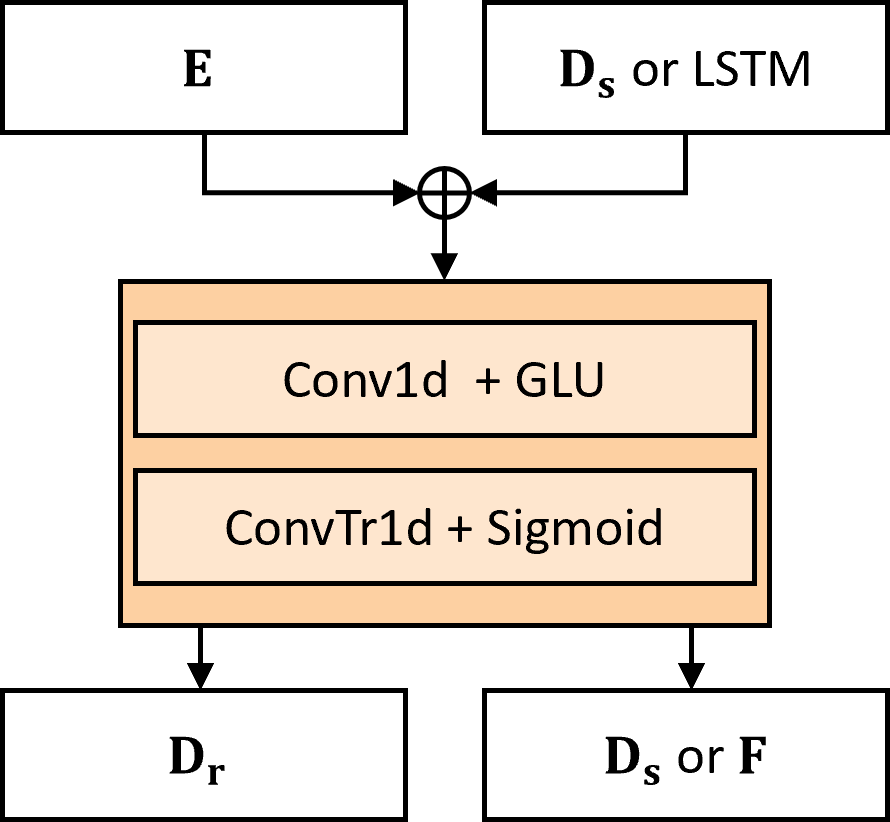}
    \caption{}
    \label{fig:suppression_block}
  \end{subfigure}
  \begin{subfigure}[b]{0.3\linewidth}
    \centering
    \includegraphics[width=\textwidth]{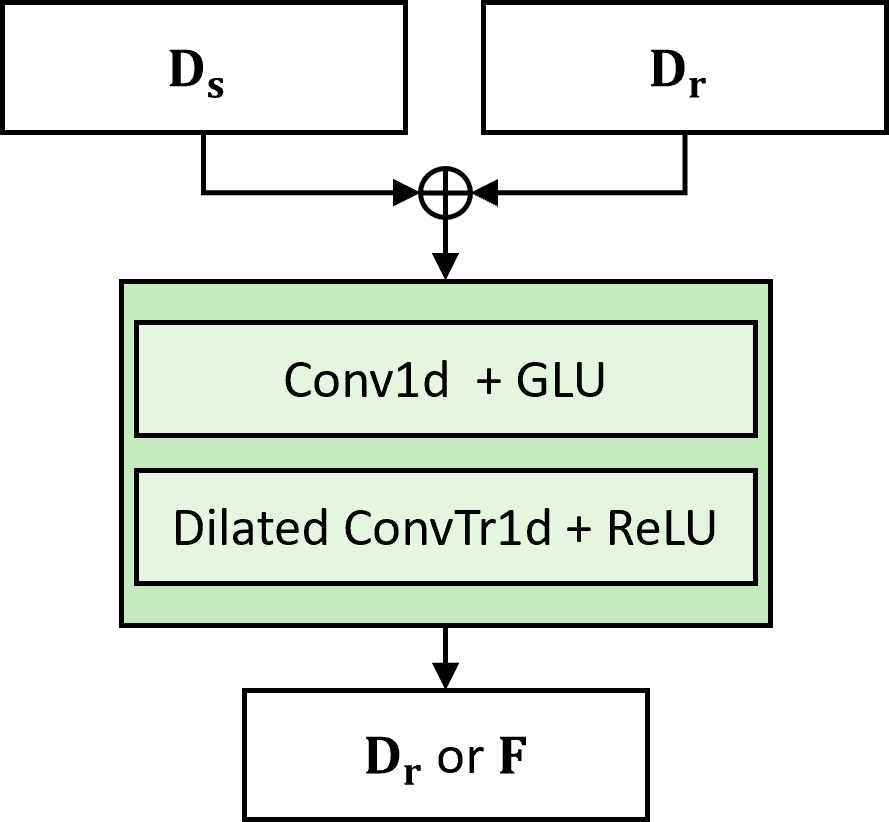}
    \caption{}
    \label{fig:generation_block} 
  \end{subfigure}
  \begin{subfigure}[b]{0.3\linewidth}
    \centering
    \includegraphics[width=\textwidth]{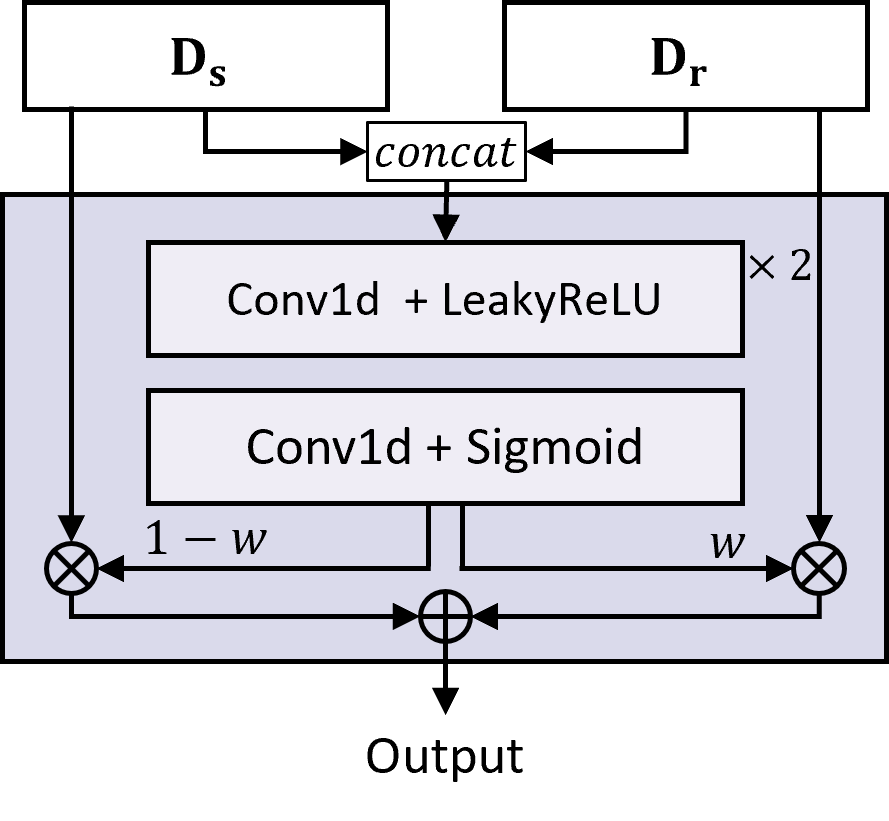}
    \caption{}
    \label{fig:fusion_block}
  \end{subfigure}
      \vspace{-5pt}
    \caption{Detailed structures of the modular blocks in \shortname: (a) Suppression; (b) Generation; (c) Fusion. }
    \label{fig:blocks}
    \vspace{-10pt}
\end{figure}

%% file: tables/baselines.tex
\begin{table*}[]
   \footnotesize
  \caption{Objective measurements of speech restoration performance on $\mathcal{A}$ testset. * indicates re-implementation.}
  \vspace{-5pt}
  \label{tab:baselines}
  \centering
  \begin{tabular}{l ||c |c c c c c c c c }
    \toprule
         \textbf{Methods}  & \multicolumn{1}{c|}{\bf \# of Params.}
     &  \multicolumn{1}{c}{\bf WV-MOS}
     & \multicolumn{1}{c}{\bf PESQ} 
     & \multicolumn{1}{c}{\bf ESTOI}
     
     &  \multicolumn{1}{c}{\bf COVL}
     &  \multicolumn{1}{c}{\bf CSIG}
     &  \multicolumn{1}{c}{\bf CBAK}
     & \multicolumn{1}{c}{\bf SRMR}  
     &  \multicolumn{1}{c}{\bf SI-SDR}
      \\ 
\midrule\midrule
Noisy   & - & 2.042 & 1.808  & 0.648 & 2.148 & 2.615 & 1.806 &  -7.282 & - \\

DEMUCS48$^*$ & 18M & 3.911 & 2.216 & 0.774  & 2.894 & 2.565 & 2.640 & 8.470 
& 5.793\\ 
DEMUCS64$^*$ & 33M & 3.842 & 2.310 & 0.782 & 2.998 & 3.696 & 2.699 & 8.633  & 6.022 \\
VoiceFixer & 122M &3.954 & 2.122 & 0.685  & 2.554 & 3.095 & 2.167 & \bf 9.301 & -20.268 \\
MetricGAN+ & 2.7M  & 2.963 & 2.379 & 0.669 & 2.551 & 2.852 & 2.254 & 8.258 & -6.743 \\
\midrule
\bf HD-DEMUCS & 24M & \bf 4.205 & \bf 2.393  & \bf 0.792 & \bf 3.067 & \bf 3.747 & \bf 2.740  & 8.999 & \bf 6.243 \\ 
    \bottomrule
  \end{tabular}
  \vspace{-10pt}
\end{table*}

%% file: 5_experiments.tex
\input{FigTex/various_test}

\section{Experiments}
\label{section:experiments}
\subsection{Experimental settings}

\paragraph{Datasets}
We utilized the Valentini dataset~\cite{valentinibotinhao16_interspeech}, consisting of the VCTK corpus with 28 English speakers~\cite{yamagishi2019cstr} and the DEMAND noise dataset~\cite{thiemann2013diverse}. 
Consistent with~\cite{valentinibotinhao16_interspeech}, we reserved one male and one female speaker, which were not included in the training set, and five distinct, unseen background noises for the test set.
The signal-to-noise ratio (SNR) was randomly selected between (0, 5, 10, 15) dB for the training set and (2.5, 7.5, 12.5, 17.5) dB for the test set.
For the training and test sets, we simulated reverberations using 243 and 27 types of room impulse responses, respectively, from the MIT Impulse Response Survey dataset~\cite{traer2016statistics}.
We simulated the spectral distortions on input speech using a low pass, high pass, and band pass filter, and frequency drop by randomly selecting a type of filter within Butterworth, Bessel, and elliptic types.
The cut-off frequencies of the low and high pass filters were randomly selected from the range of 4k-7.5kHz and 10-100 Hz, respectively and use the same range of cut-off frequencies for the bandpass filters.
For the bandlimited test sets, only the low pass filters were applied, where cut-off frequencies are set uniformly in (4, 5, 6, 7) kHz.
We constructed 4 different subsets to exhibit the effectiveness of our model on each distortion: $\mathcal{N}$ (noisy speech), $\mathcal{R}$ (noisy and reverberant speech), $B$ (band-limited speech), and $\mathcal{A}$ (speech with all distortions).
\input{FigTex/decoder_samples}

\paragraph{Evaluation metrics}
We assessed the performance of the speech restoration task using various metrics.
The speech quality is evaluated with the wide-band Perceptual Evaluation of Speech Quality (PESQ)~\cite{union2007wideband}, while the speech intelligibility was measured in extended Short-Time Objective Intelligibility (ESTOI)~\cite{jensen2016modulated}.
To evaluate speech dereverberation performance, we used the Speech-to-Reverberation Modulation Energy Ratio (SRMR) metric \cite{falk2010temporal}.
For the waveform reconstruction, we measured the scale-invariant signal-to-distortion ratio (SI-SDR)~\cite{le2019sdr}, while the restoration performance, including bandwidth extension, was evaluated using the Wideband Voice Mean-Opinion-Score~(WV-MOS)~\cite{andreev2022hifi++}.
We also utilized composite measurements to analyze the overall quality (COVL), signal distortion (CSIG), and background noise (CBAK)~\cite{yi2008evaluation}.
A higher score on all evaluation metrics indicates an improved performance.

\paragraph{Training configuration}
We trained the encoder-decoder network first, then added a fusion block for joint training. 
Before attaching the fusion block, the outputs were combined with a 0.5 weight value. 
We used Adam optimizer~\cite{kingma2014adam} with a learning rate of 0.0003, a cosine annealing scheduler for training.

\subsection{Results}

\paragraph{Comparison with baselines}
We re-implemented two baseline models, DEMUCS48 and DEMUCS64, using 48 and 64 hidden channels, for the fair performance comparison by training in an identical environmental setting with the proposed model.
Additionally, we brought the pre-trained parameters of VoiceFixer~\cite{liu2022voicefixer} and MetricGAN+~\cite{fu2021metricgan+} for comparisons.
\Table{baselines} displays the experimental results of comparing the baseline models with the $\mathcal{A}$ test set.
The proposed model outperformed the baseline models in terms of speech quality (PESQ), intelligibility (ESTOI), and, particularly speech restoration (WV-MOS).
Furthermore, the composite metrics and SI-SDR results demonstrate the effectiveness of \shortname~in the suppression task.
We conducted a detailed analysis by comparing the scores of input, DEMUCS48, DEMUCS64, and \shortname~across subsets in \Figure{subset}.
For the SRMR metric, we evaluated against $\mathcal{R}$ and $\mathcal{A}$ subsets, which contain reverberation distortions.
These demonstrate the proposed model exhibits robustness across various distortions and that it offers superior performances in harsh conditions such as test sets $\mathcal{R}$ and $\mathcal{A}$.

\input{tables/ablation_decoders}

\paragraph{Analysis of decoders}
In \Table{ablation_decoders}, we report the quality of output waveforms of each decoder to investigate their individual contributions to the proposed model. 
The results demonstrate that the superior performance of the model is attributed to the \refinement~module. 
The `\suppression~output’ result indicates the suppression decoder does not produce high-quality speech signals but exhibits its ability to suppress background noise in terms of the CBAK metric.
Figure 4 supports the findings of \Table{ablation_decoders} by presenting the spectrograms of the reference, distorted input, and outputs of the decoders and \shortname.
The figures clearly demonstrate that each decoder performs properly for its designed restoration task without additional training loss to each module.
Consistent with the `\suppression~output' results, Figure 4(d) confirms that the poor quality of the suppression decoder results from the over-suppression issue associated with powerful suppression of various distortions.

\input{tables/ablation_modules}

\paragraph{Ablation studies}
To investigate the contribution of each module in \shortname, we trained several models with specific modules selectively removed, and the results are presented in \Table{table_ablation_composite}.
``w/o $\mathbf{F}$" model summed up the outputs of both decoders without learnable weights of the fusion block.
``w/o $\mathbf{F, S_{D_s \rightarrow D_r}}$" model removed the skip connection between the two decoders, but kept the skip connections $\mathbf{S_{E \rightarrow D_s}}$. 
``w/o $\mathbf{F, D_r}$" model used only the suppression decoder with the skip connections $\mathbf{S_{E \rightarrow D_s}}$.
``w/o $\mathbf{F, D_s}$" used only the refinement decoder with the skip connections $\mathbf{S_{E \rightarrow D_r}}$.
The fusion block \fusion, aggregating the outputs of heterogeneous decoders using learnable weights, improves overall performance compared to using a fixed weight of 0.5.
Moreover, the absence of \skipconnect~revealed a noticeable drop in speech quality compared to the suppression performance, confirming the importance of the enhanced features to the refinement decoder.
Notably, the removal of the refinement decoder \refinement~led to significant performance degradation compared to other models, highlighting its effectiveness in high-quality speech restoration with various distortion present.
On the other hand, while there was minor performance degradation in the absence of the suppression decoder \suppression, it still demonstrated its capability to suppress distortions when it was used solely, without the refinement decoder.

%% file: FigTex/various_test.tex
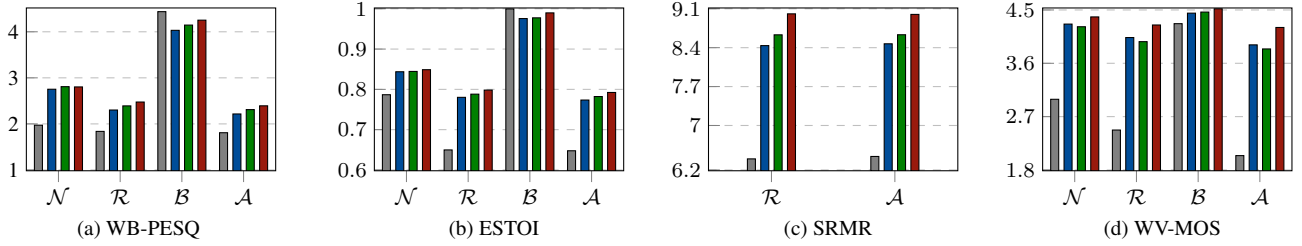
\begin{figure*}[t]
   \footnotesize
    \begin{minipage}[t]{0.22\linewidth}
        \centering
        \begin{tikzpicture}
        \begin{axis}[
        width=1.3\columnwidth, height=\linewidth,
        scaled y ticks = false,
        scaled x ticks = false,
        xtick pos=left, ytick pos=left,
        ybar,
        bar width=3pt,
        enlargelimits=0.005,
        xmin=-.5, xmax=3.5,
        ymin=1,   ymax=4.5,
        xticklabels={$\mathcal{N}$, $\mathcal{R}$, $\mathcal{B}$, $\mathcal{A}$},
        xtick={0, 1, 2, 3, 4, 5},
        ytick={0, 1, 2, 3, 4, 5},
        yticklabels={0, 1, 2, 3, 4, 5},
        minor tick length=1ex,
        major x tick style = {opacity=1},
        ymajorgrids=true,
        xmajorgrids=false,
        grid style=dashed,
        legend style={at={(0.7497, 1.8)},
        anchor=north, legend columns=-3},
        ]
        \addplot[black, fill=gray] coordinates{
        (0, 1.9720) (1,1.8372) (2,4.4372) (3,1.8083) }; 
        \addplot[black, fill=blueish] coordinates{
        (0, 2.7527) (1,2.3013) (2,4.0318) (3,2.2157) }; 
        \addplot[black, fill=dark_green] coordinates{
        (0, 2.8084) (1,2.3910) (2,4.1458) (3,2.3103) }; 
        \addplot[black, fill=darkred] coordinates{
        (0, 2.8027) (1,2.4749) (2,4.2526) (3,2.3925) }; 
        \end{axis}
        \end{tikzpicture}
        \centerline{(a) WB-PESQ}
        \vspace{-5pt}
    \end{minipage}
    \hfill
    \begin{minipage}[t]{0.22\linewidth}
    \centering
        \begin{tikzpicture}
        \begin{axis}[
        width=1.3\columnwidth, height=\linewidth,
        scaled y ticks = false,
        scaled x ticks = false,
        xtick pos=left, ytick pos=left,
        ybar,
        bar width=3pt,
        enlargelimits=0.005,
        xmin=-.5, xmax=3.5,
        ymin=0.6,   ymax=1,
        xticklabels={$\mathcal{N}$, $\mathcal{R}$, $\mathcal{B}$, $\mathcal{A}$},
        xtick={0, 1, 2, 3},
        ytick={0.6, 0.7, 0.8, 0.9, 1},
        minor tick length=1ex,
        major x tick style = {opacity=1},
        ymajorgrids=true,
        xmajorgrids=false,
        grid style=dashed,
        legend style={at={(0.7497, 1.8)},
        anchor=north, legend columns=-3},
        ]
        \addplot[black, fill=gray] coordinates{
        (0, 0.7867) (1,0.6497) (2,0.9997) (3,0.6476) };  
        \addplot[black, fill=blueish] coordinates{
        (0, 0.8437) (1,0.7802) (2,0.9757) (3,0.7735) }; 
        \addplot[black, fill=dark_green] coordinates{
        (0, 0.8445) (1,0.7881) (2,0.9775) (3,0.7822) }; 
        \addplot[black, fill=darkred] coordinates{
        (0, 0.8488) (1,0.7982) (2,0.9898) (3,0.7923) }; 
        
        \end{axis}
        \end{tikzpicture}
        \centerline{\qquad(b) ESTOI}
        \vspace{-5pt}
    \end{minipage}
    \hfill
    \begin{minipage}[t]{0.22\linewidth}
    \centering
        \begin{tikzpicture}
        \begin{axis}[
        width=1.3\columnwidth, height=\linewidth,
        scaled y ticks = false,
        scaled x ticks = false,
        xtick pos=left, ytick pos=left,
        ybar,
        bar width=3pt,
        enlargelimits=0.005,
        xmin=-0.5, xmax=1.5,
        ymin=6.2,   ymax=9.1,
        xticklabels={$\mathcal{R}$, $\mathcal{A}$},
        xtick={0, 1},
        ytick={6.2, 7, 7.7, 8.4, 9.1},
        minor tick length=1ex,
        major x tick style = {opacity=1},
        ymajorgrids=true,
        xmajorgrids=false,
        grid style=dashed,
        legend style={at={(0.7497, 1.8)},
        anchor=north, legend columns=-3},
        ]
        \addplot[black, fill=gray] coordinates{ 
        (0,6.3991) (1,6.4428) }; 
        \addplot[black, fill=blueish] coordinates{
        (0,8.4371) (1,8.4698) }; 
        \addplot[black, fill=dark_green] coordinates{
        (0,8.6312) (1,8.6327) }; 
        \addplot[black, fill=darkred] coordinates{
        (0,9.0086) (1,8.9987) }; 
        
        
        \end{axis}
        \end{tikzpicture}
        \centerline{\qquad(c) SRMR}
        \vspace{-5pt}
    \end{minipage}
    \hfill
    \begin{minipage}[t]{0.22\linewidth}
    \centering
        \begin{tikzpicture}
        \begin{axis}[
        width=1.3\columnwidth, height=\linewidth,
        scaled y ticks = false,
        scaled x ticks = false,
        xtick pos=left, ytick pos=left,
        ybar,
        bar width=3pt,
        enlargelimits=0.005,
        xmin=-.5, xmax=3.5,
        ymin=1.8,   ymax=4.52,
        xticklabels={$\mathcal{N}$, $\mathcal{R}$, $\mathcal{B}$, $\mathcal{A}$},
        xtick={0, 1, 2, 3},
        ytick={1.8, 2.7, 3.6, 4.5},
        minor tick length=1ex,
        major x tick style = {opacity=1},
        ymajorgrids=true,
        xmajorgrids=false,
        grid style=dashed,
        legend style={at={(0.7497, 1.8)},
        anchor=north, legend columns=-3},
        ]
        \addplot[black, fill=gray] coordinates{
        (0, 2.9925) (1,2.4736) (2,4.2686) (3,2.0420) }; 
        \addplot[black, fill=blueish] coordinates{
        (0, 4.2616) (1,4.0346) (2,4.4461) (3,3.9109) }; 
        \addplot[black, fill=dark_green] coordinates{
        (0, 4.2169) (1,3.9651) (2,4.4645) (3,3.8421) }; 
        \addplot[black, fill=darkred] coordinates{
        (0, 4.3826) (1,4.2464) (2,4.5198) (3,4.2045) }; 
        
        \end{axis}
        \end{tikzpicture}
        \centerline{\qquad(d) WV-MOS}
        \vspace{-5pt}
    \end{minipage}
    \vspace{-10pt}
    \caption{Objective measurements on various distortion test sets of input, baseline, and proposed models. $\mathcal{N}$, $\mathcal{R}$, $\mathcal{B}$, $\mathcal{A}$ indicate tests sets for noisy, noisy-reverberant, bandlimited, and all three distortions, respectively. \tgray{Gray}, \tblue{Blue}, \tgreen{Green}, and \tred{Red} bars represent distorted inputs, DEMUCS48 outputs, DEMUCS64 outputs, and HD-DEMUCS outputs, respectively.}
    \label{fig:subset}
    \vspace{-15pt}
\end{figure*}

%% file: FigTex/decoder_samples.tex
\begin{figure}[t]
\footnotesize
\vspace{-5 pt}
\centering
 \begin{tikzpicture}    
 \matrix (fig) [matrix of nodes]{
 \includegraphics[width=0.17\columnwidth]{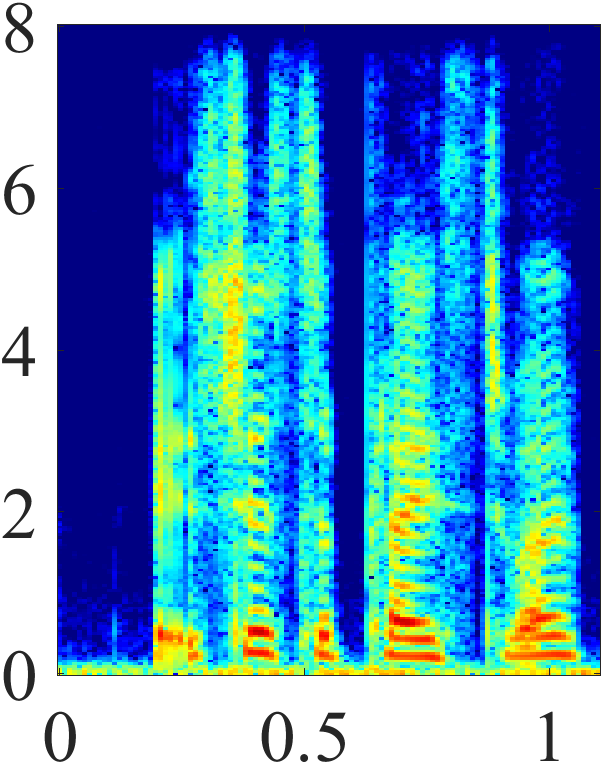}
 &
 \includegraphics[width=0.17\columnwidth]{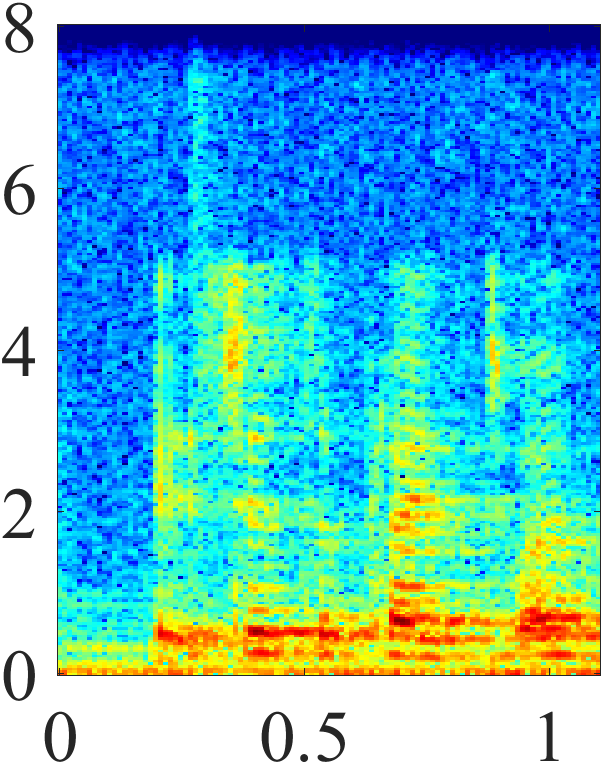}
 &
 \includegraphics[width=0.17\columnwidth]{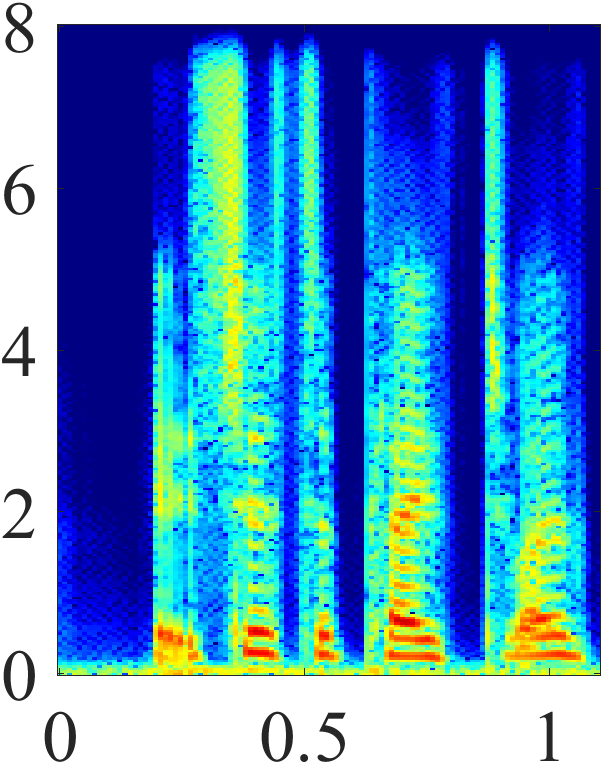}
 &
 \includegraphics[width=0.17\columnwidth]{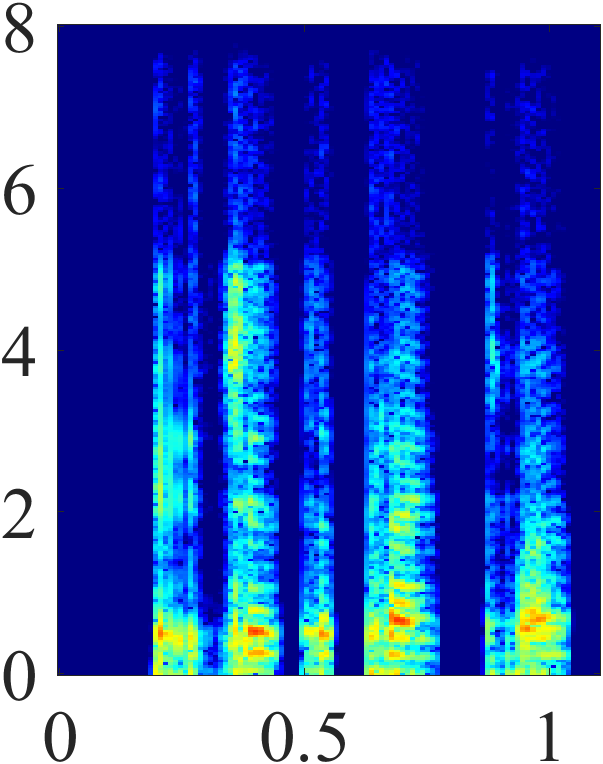}
 &
 \includegraphics[width=0.17\columnwidth]{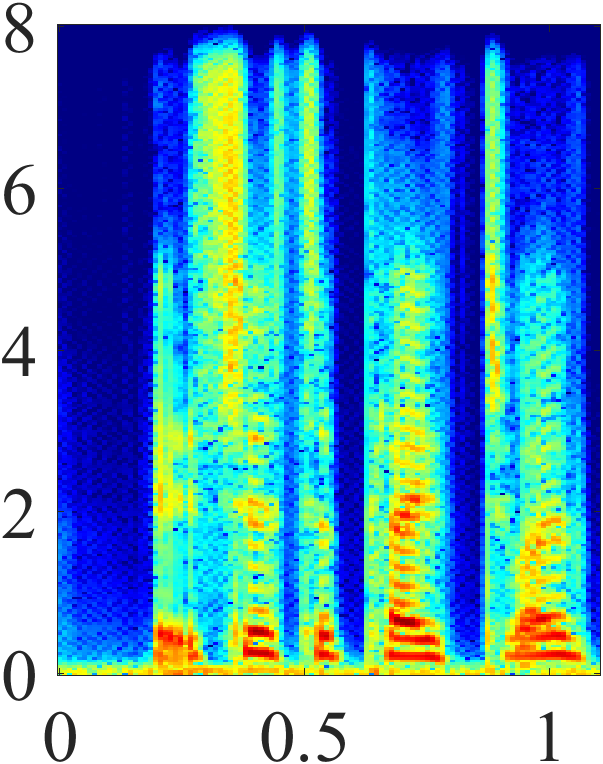}
 \\
 |[text width=0.17\columnwidth]| {\subcaption{}}
 &
 |[text width=0.17\columnwidth]| {\subcaption{}}
 &
 |[text width=0.17\columnwidth]| {\subcaption{}}
 &
 |[text width=0.17\columnwidth]| {\subcaption{}}
 &
 |[text width=0.17\columnwidth]| {\subcaption{}}
 \\
  };
 \path([xshift=-1mm, yshift=1mm]fig-1-1.west)  -- ([xshift=-1mm, yshift=1mm]fig-1-1.west) node[midway,sloped]{\tiny{Frequency (kHz)}}; 
 \path ([yshift=-1.5mm]fig-1-1.south)  -- ([yshift=-1.5mm]fig-1-5.south) node[midway]{\tiny{Time (s)}}; 
 \end{tikzpicture}
 \vspace{-25pt}
 \caption{Qualitative results. Spectrogram of (a) clean speech, (b) input speech, (c) \shortname~output, (d) \suppression~output, and (e) \refinement~output}
 \vspace{-5pt}
\label{fig:samples}
\end{figure}

%% file: tables/ablation_decoders.tex
\begin{table}[t]
\footnotesize
\centering
\caption{Analysis of the each HD-DEMUCS decoders outputs on $\mathcal{A}$ testset.}
\vspace{-10pt}
\setlength\tabcolsep{4 pt}
\begin{tabular}{c || c c c c c}
\toprule
      \multicolumn{1}{c||}{\bf Methods}
      &  \multicolumn{1}{c}{\bf WV-MOS} & \multicolumn{1}{c}{\bf PESQ} 
      & \multicolumn{1}{c}{\bf COVL}
      & \multicolumn{1}{c}{\bf CSIG} & \multicolumn{1}{c}{\bf CBAK} \\ 
\midrule 
\multicolumn{1}{l||} {Noisy}  & 2.042 & 1.808 & 2.148 & 2.615  & 1.806 \\
\multicolumn{1}{l||} {\bf HD-DEMUCS}  & \bf 4.205 & \bf 2.393 & \bf 3.067 & \bf 3.747  & \bf 2.740 \\
\midrule
\multicolumn{1}{l||}{\refinement~output} & 4.198 & 2.279 & 3.035 & 3.719 & 2.486  \\
\multicolumn{1}{l||}{\suppression~output} & 2.536 & 1.428 & 1.497 & 1.666 & 2.040 \\

\bottomrule
\end{tabular}
\label{tab:ablation_decoders}
\vspace{-20pt}
\end{table}

%% file: tables/ablation_modules.tex
\begin{table}[t]
\footnotesize
\centering
\caption{Ablation study for the strategy in the proposed model on $\mathcal{A}$ testset.}
\vspace{-10pt}
\setlength\tabcolsep{4 pt}

\begin{tabular}{c || c c c c c}
\toprule
      \multicolumn{1}{c||}{\bf Methods}
      &  \multicolumn{1}{c}{\bf WV-MOS} & \multicolumn{1}{c}{\bf PESQ} 
      & \multicolumn{1}{c}{\bf COVL}
      & \multicolumn{1}{c}{\bf CSIG} & \multicolumn{1}{c}{\bf CBAK} \\ 
\midrule 
\multicolumn{1}{l||} {Noisy}  & 2.042 & 1.808 & 2.148 & 2.615  & 1.806 \\
\multicolumn{1}{l||} {\bf HD-DEMUCS}  & \bf 4.205 & \bf 2.393 & \bf 3.067 & \bf 3.747  & \bf 2.740 \\
\midrule

\multicolumn{1}{l||}{{\scriptsize w/o} \fusion} & 4.167 & 2.379 & 3.052 & 3.731 & 2.726 \\

\multicolumn{1}{l||}{{\scriptsize w/o} \fusion, \skipconnect} & 3.985 & 2.188 & 2.867 & 3.566 & 2.606 \\

\multicolumn{1}{l||}{{\scriptsize w/o} \fusion, \refinement} & 3.281 & 1.510 & 2.435 & 3.154 & 2.222 \\
\multicolumn{1}{l||}{{\scriptsize w/o} \fusion, \suppression} & 4.111 & 2.235 & 3.034 & 3.719 & 2.710 \\
\bottomrule
\end{tabular}
\label{tab:table_ablation_composite}
\vspace{-20pt}
\end{table}

%% file: 6_conclusion.tex
\section{Conclusions}
In this paper, we proposed an end-to-end speech restoration model, \fullname~(\shortname), that utilizes two heterogeneous decoders for two different perspectives of restoration: suppression and refinement.
\shortname~demonstrated powerful suppression performance through a mask estimation approach of suppression decoder and the effectiveness of refinement decoder with dilated convolution layers.
Additionally, we incorporated a fusion block to combine effectively the outputs of the two decoders by predicting a learnable weighting value.
We evaluated the proposed model and baselines in the presence of various distortions with objective measurements, demonstrating the superiority of \shortname.
Specifically, we analyzed the contributions of each module in \shortname~with ablation studies and qualitative results and confirmed the intended functionality of each module.
Further improvements could be achieved using a larger dataset or by incorporating speech-related features such as pitch during training.